\PassOptionsToPackage{dvips}{graphicx}
\documentclass[english]{cccconf}

\usepackage{type1cm}
\usepackage{cite}      

\usepackage{psfrag}    

\usepackage{amsmath,amssymb}   

\begin{document}

\title{Kalman Filtering in the Presence of State Space Equality Constraints}
\author{Nachi Gupta}

%

\affiliation{Oxford University Computing Laboratory, Numerical Analysis Group, Wolfson Building, Parks Road, Oxford OX1 3QD, U.K.\email{nachi@comlab.ox.ac.uk}}

\maketitle

\begin{abstract}
We discuss two separate techniques for Kalman Filtering in the presence of state space equality constraints.  We then prove that despite the lack of similarity in their formulations, under certain conditions, the two methods result in mathematically equivalent constrained estimate structures.  We conclude that the potential benefits of using equality constraints in Kalman Filtering often outweigh the computational costs, and as such, equality constraints, when present, should be enforced by way of one of these two methods.
\end{abstract}


\footnotetext{The author would like to thank Raphael Hauser for reading drafts of this paper and the Clarendon Bursary for financial support.}

\section{Introduction}
%
%
%
%

Kalman Filtering \cite{Kalman1960} is a method to make real-time predictions for systems with some known dynamics.  Traditionally, problems requiring Kalman Filtering have been complex and nonlinear.  Many advances have been made in the direction of dealing with nonlinearities (e.g., Extended Kalman Filter \cite{BLK2001}, Unscented Kalman Filter \cite{JU1997}). These problems also tend to have inherent state space {\em equality} constraints (e.g., a fixed speed for a robotic arm) or even state space {\em inequality} constraints (e.g., maximum attainable speed of a motor).  In the past, less interest has been generated towards constrained Kalman Filtering, partly because constraints can be difficult to model.  As a result, equality constraints are often neglected in standard Kalman Filtering applications.  However, the benefits of incorporating constraints can outweigh the computational costs associated with constraining the estimate (e.g., the constrained estimate can be quite different from the unconstrained estimate and the error covariance matrix can only get tighter since we are adding information to our model).  

We discuss two distinct approaches to generalizing an equality constrained Kalman Filter.  The first approach is to augment the measurement space of the filter with the equality constraints (i.e., as perfect noise-free measurements) at each iteration.  The second approach is to find the unconstrained estimate from a Kalman Filter and project it down to the equality constrained space.  Both of these approaches have appeared in the literature in the past (e.g., \cite{TS1990}, \cite{SC2002}).  We will then show that, under certain conditions, the first approach and the second approach actually yield the same analytical distribution for the constrained estimate despite the differing formulations.  There is a third well-known approach to this problem, which is to reduce the state space by the dimension of the constraints.  This can lead to a state space that does not carry much meaning to the engineer.  This approach, while valid, is not discussed in this paper.

Analogous to the way a Kalman Filter can be extended to solve problems containing non-linearities in the dynamics using an Extended Kalman Filter by linearizing locally (or by using an Unscented Kalman Filter), linear equality constrained filtering can similarly be extended to problems with nonlinear constraints by linearizing locally (or by way of another scheme).  The accuracy achieved by methods dealing with nonlinear constraints will naturally depend on the structure and curvature of the nonlinear function itself.

Equality constrained Kalman Filtering also appears as a subroutine in the more general framework of inequality constrained Kalman Filtering.  One method for extending an equality constrained filter to an inequality constrained filter would be to use an active set method (as in \cite{GHJ2005}).

\section{Kalman Filter} \label{sec::kf}

A discrete-time Kalman Filter attempts to find the best running estimate for a recursive system governed by the following model:

\begin{equation} \label{kfsm} 
x_{k} = F_{k,k-1} x_{k-1} + u_{k,k-1}, \qquad u_{k,k-1} \sim N(0,Q_{k,k-1}) 
\end{equation}

\begin{equation} \label{kfmm} 
z_{k} = H_{k} x_{k} + v_{k}, \qquad v_{k} \sim N(0,R_{k}) 
\end{equation}

Here $x_{k}$ represents the true state of the underlying system\footnote{The subscript $k$ means for the $k$-th time step.} and $F_{k,k-1}$ is the matrix that describes the transition dynamics of the system from  $x_{k-1}$ to $x_{k}$.  The measurement made by the observer is denoted $z_{k}$, and $H_{k}$ is the matrix that transforms a vector from the state space into the appropriate vector in the measurement space.  The noise terms $u_{k,k-1}$ and $v_{k}$ encompass known and unknown errors in $F_{k,k-1}$ and $H_{k}$ and are normally distributed with mean 0 and variances $Q_{k,k-1}$ and $R_{k}$, respectively.  At each iteration, the Kalman Filter makes a state prediction for $x_k$, which we denote by $\hat{x}_{k|k-1}$.  We use the notation ${k|k-1}$ since we will only use measurements provided until time-step $k-1$ in order to make the prediction at time-step $k$.  The state prediction error $\tilde{x}_{k|k-1}$ is defined as the difference between the true state and the state prediction, as below.

\begin{equation} \label{se1}
\tilde{x}_{k|k-1} = x_{k} - \hat{x}_{k|k-1}
\end{equation}

The covariance structure for the expected error on the state prediction is defined as the expectation of the outer product of the state prediction error.  We call this covariance structure the error covariance prediction and denote it $P_{k|k-1}$.

\begin{equation} \label{P-outer1}
P_{k|k-1} = \mathbb{E}\left[\left(\tilde{x}_{k|k-1}\right)\left(\tilde{x}_{k|k-1}\right)'\right]
\end{equation}

In addition, the filter will provide a state estimate for $x_{k}$, given all the measurements provided up to and including time step $k$.  We denote these estimates by $\hat{x}_{k|k}$.  We similarly define the state estimate error $\tilde{x}_{k|k}$ as below.

\begin{equation} \label{se2}
\tilde{x}_{k|k} = x_{k} - \hat{x}_{k|k}
\end{equation}

The expectation of the outer product of the state estimate error represents the covariance structure of the expected errors on the state estimate, which we call the updated error covariance and denote $P_{k|k}$.

\begin{equation} \label{P-outer2}
P_{k|k} = \mathbb{E}\left[\left(\tilde{x}_{k|k}\right)\left(\tilde{x}_{k|k}\right)'\right]
\end{equation}

At time-step $k$, we can make a prediction for the underlying state of the system by allowing the state to transition forward using our model for the dynamics and noting that $\mathbb{E}\left[u_{k,k-1}\right] = 0$.  This serves as our state prediction.

\begin{equation} \label{kfsp} 
\hat{x}_{k|k-1} = F_{k,k-1} \hat{x}_{k-1|k-1} 
\end{equation}

If we expand the expectation in Equation \eqref{P-outer1}, we have the following equation for the error covariance prediction.\footnote{We use the prime notation on a vector or a matrix to denote its transpose throughout this paper.}

\begin{equation} \label{kfcp} 
P_{k|k-1} = F_{k,k-1} P_{k-1|k-1} F_{k,k-1}' + Q_{k,k-1}
\end{equation}

We can transform our state prediction into the measurement space, which is a prediction for the measurement we now expect to observe.

\begin{equation} \label{kfmp} 
\hat{z}_{k|k-1} = H_{k} \hat{x}_{k|k-1}
\end{equation}

The difference between the observed measurement and our predicted measurement is the measurement residual, which we are hoping to minimize in this algorithm.

\begin{equation} \label{kfi} 
\nu_{k} = z_{k} - \hat{z}_{k|k-1} 
\end{equation}

We can also calculate the associated covariance for the measurement residual, which is the expectation of the outer product of the measurement residual with itself, $\mathbb{E}\left[\nu_k \nu_k'\right]$.  We call this the measurement residual covariance.

\begin{equation} \label{kfic} 
S_{k} = H_{k} P_{k|k-1} H_{k}' + R_{k} 
\end{equation}

We now calculate the Kalman Gain, which lies at the heart of the Kalman Filter.  This tells us how much we prefer our new observed measurement over our state prediction. 

\begin{equation} \label{kfkg} 
K_{k} = P_{k|k-1} H_{k}' S_{k}^{-1} 
\end{equation}

Using the Kalman Gain and measurement residual, we update the state estimate.  If we look carefully at the following equation, we are taking a weighted sum of our state prediction with the Kalman Gain multiplied by the measurement residual, so the Kalman Gain is telling us how much to `weigh in' information contained in the new measurement.  We calculate the updated state estimate by

\begin{equation} \label{kfsu} 
\hat{x}_{k|k} = \hat{x}_{k|k-1} + K_{k}  \nu_{k}
\end{equation}

Finally, we calculate the updated error covariance by expanding the outer product in Equation \eqref{P-outer2}.\footnote{The $I$ in Equation \eqref{kfcu} represents the identity matrix of the appropriate dimension.  Throughout the remainder of this paper, we will continue to use $I$ in the same fashion.}

\begin{equation} \label{kfcu} 
P_{k|k} = \left(I - K_{k} H_{k}\right) P_{k|k-1}  
\end{equation}

The covariance matrices in the Kalman Filter provide us with a measure for uncertainty in our predictions and updated state estimate.  This is a very important feature for the various applications of filtering since we then know how much to trust our predictions and estimates.  Also, since the method is recursive, we need to provide an initial covariance that is large enough to contain the initial state estimate to ensure comprehendible performance.  For a more detailed discussion of Kalman Filtering, we refer the reader to the following book \cite{BLK2001}.

\section{Incorporating Equality Constraints by Augmenting the Measurement Space} \label{sec::ams}

The first method for incorporating equality constraints into a Kalman Filter is to `observe' the constraints at every iteration as noise-free measurements.  To illustrate this, we augment linear constraints to the system shown in Equations \eqref{kfsm} and \eqref{kfmm} as measurements with 0 variance.  We will define the constraints in this formulation as $D_k x_k =\delta_k $.\footnote{\label{fn::D}We assume these constraints are well defined throughout this paper -- i.e., no constraints conflict with one another to cause a null solution and no constraints are repeated.  More specifically, we assume $D_k$ has full row rank.  Note that under these conditions if $D_k$ was a square matrix, the constraints would completely determine the state.}  Thus, we can re-write the system.

\begin{equation} x_{k}^D = F_{k,k-1} x^D_{k-1} + u_{k,k-1}, \qquad u_{k,k-1} \sim N(0,Q_{k,k-1}) \end{equation}

\begin{equation} z_{k}^D = H^D_{k} x^D_{k} + v^D_{k}, \qquad v_{k} \sim N(0,R^D_{k}) \end{equation}

Here we use the superscript $D$ notation to denote the new filter with the equality constraints.  The next three equations show the construction of the augmentation in the measurement space.

\begin{equation}
z_k^D = 
\begin{bmatrix} 
	z_{k} \\
	\delta_k
\end{bmatrix}
\end{equation}

\begin{equation} \label{HkD}
H_k^D= 
\begin{bmatrix}
	H_{k} \\
	D_k 
\end{bmatrix}
\end{equation}

\begin{equation} \label{RD}
R_k^D=
\begin{bmatrix}
	R_{k} & 0 \\
	0 & 0
\end{bmatrix}
\end{equation}

The augmented state now forces $D_k x_k^D$ to be equal to $\delta_k$ exactly (i.e., with no noise term) at every iteration.\footnote{With $x_k^D$ is constructed in the same fashion as $x_k$.}  Let us now expand the equations for the Kalman Filter prediction and update to gain a stronger understanding of how the filter has changed.  

The state prediction from Equation \eqref{kfsp} becomes the following.

\begin{equation} \label{kfspd}
\hat{x}^D_{k|k-1} = F_{k,k-1} \hat{x}^D_{k-1|k-1}
\end{equation}

The error covariance prediction from Equation \eqref{kfcp} becomes the following.

\begin{equation} \label{kfcpd}
P^D_{k|k-1} = F_{k,k-1} P^D_{k-1|k-1} F_{k,k-1}' + Q_{k,k-1}
\end{equation}

The measurement prediction from Equation \eqref{kfmp} can then be written in the following form.

\begin{subequations} \label{kfmpd}
\begin{align} 
\hat{z}^D_{k|k-1} & = H^D_{k} \hat{x}^D_{k|k-1} \\
& =
	\begin{bmatrix}
		H_{k} \hat{x}^D_{k|k-1}\\
		D_k \hat{x}^D_{k|k-1}
	\end{bmatrix}
\end{align}
\end{subequations}

Similarly, we can express the measurement residual from Equation \eqref{kfi} in the following manner.

\begin{subequations} \label{kfid}
\begin{align} 
\nu^D_{k} & = z_{k}^D - \hat{z}^D_{k|k-1} \\
& =
	\begin{bmatrix}
		z_k - H_{k} \hat{x}^D_{k|k-1}\\
		\delta_k -D_k \hat{x}^D_{k|k-1}
	\end{bmatrix}
\end{align}
\end{subequations}

We expand the measurement residual covariance from Equation \eqref{kfic} below.

\begin{subequations} \label{kficd}
\begin{align} 
S^D_{k} & = H^D_{k} P^D_{k|k-1} \left(H^{D}_{k}\right)' + R^D_{k} \\
& = 
	\begin{bmatrix}
		H_{k} \\
		D_k 
	\end{bmatrix}
	P^D_{k|k-1}
	\begin{bmatrix}
		H_k' & D_k' 
	\end{bmatrix}
	+
	\begin{bmatrix}
		R_k & 0 \\
		0 & 0
	\end{bmatrix} \\
& = 
	\begin{bmatrix} 
		H_k P^D_{k|k-1} H_k' + R_k & H_k P^D_{k|k-1} D_k' \\
		D_k P^D_{k|k-1} H_k' & D_k P^D_{k|k-1} D_k'
	\end{bmatrix}
\end{align}
\end{subequations}

The Kalman Gain can now be written as below.

\begin{equation}
K^D_{k} = P^D_{k|k-1} \left(H^D_{k}\right)' \left(S^D_{k}\right)^{-1}
\end{equation}

In order to further expand this term, we denote $\left(S^D_k\right)^{-1}$ in the following block matrix form. 

\begin{equation} \label{Sinv}
\begin{bmatrix}
	\left(S_k^D\right)^{-1}_a & \left(S_k^D\right)^{-1}_b \\
	\left(S_k^D\right)^{-1}_c & \left(S_k^D\right)^{-1}_d
\end{bmatrix}
\end{equation}

We then expand the Kalman Gain in terms of the block structure of Equation \eqref{Sinv}.

\begin{subequations} \label{kfkgd}
\begin{align}
K_k^D & = P_{k|k-1}^D
	\begin{bmatrix}
		H_k' & D_k'
	\end{bmatrix}
	\begin{bmatrix}
		\left(S_k^D\right)^{-1}_a & \left(S_k^D\right)^{-1}_b \\
		\left(S_k^D\right)^{-1}_c & \left(S_k^D\right)^{-1}_d
	\end{bmatrix} \\
& = 
	\begin{bmatrix}
		P_{k|k-1}^D H_k' & P_{k|k-1}^D D_k' 
	\end{bmatrix}
	\begin{bmatrix}
		\left(S_k^D\right)^{-1}_a & \left(S_k^D\right)^{-1}_b \\
		\left(S_k^D\right)^{-1}_c & \left(S_k^D\right)^{-1}_d
	\end{bmatrix} \\
& = \label{kfkgd-split}
	\begin{bmatrix}
	\left(K_k^D\right)_a & \left(K_k^D\right)_b
	\end{bmatrix}
\end{align}
\end{subequations}

Here, we've used the following two terms to shorten the expression above.

\begin{subequations}
\begin{align}
\label{KDa} \left(K_k^D\right)_a & = P_{k|k-1}^D H_k' \left(S_k^D\right)^{-1}_a + P_{k|k-1}^D D_k' \left(S_k^D\right)^{-1}_c \\
\label{KDb} \left(K_k^D\right)_b & = P_{k|k-1}^D H_k' \left(S_k^D\right)^{-1}_b + P_{k|k-1}^D D_k' \left(S_k^D\right)^{-1}_d
\end{align}
\end{subequations}

Furthermore, the updated state estimate from Equation \eqref{kfsu} takes the following form.

\begin{equation}
\hat{x}^D_{k|k} = \hat{x}^D_{k|k-1} +K^D_{k}  \nu^D_{k} \label{kfsud}
\end{equation}

And the updated error covariance from Equation \eqref{kfcu} changes in the following way.

\begin{equation}
P^D_{k|k} = (I - K^D_{k} H^D_{k}) P^D_{k|k-1} \label{kfcud}
\end{equation}

Methods using augmentation in Kalman Filters have appeared for different applications in the past (e.g., Fixed-Point Smoothing \cite{GMS1988}, Bias Detection \cite{Friedland1969}).  In order to gain a stronger understanding of the effects of augmentation in Kalman Filters, it can be helpful to read and understand these methods, as well.

\subsection{Improvement gained over an Unconstrained Filter}

For a given iteration, we are interested in the improvement gained by using this method over a method that does not incorporate equality constraints.  In order to do so, we would like to find the constrained estimated $\hat{x}^D_{k|k}$ in terms of the unconstrained estimate $\hat{x}_{k|k}$ (and similarly the constrained error covariance matrix $P^D_{k|k}$ in terms of the unconstrained error covariance matrix $P_{k|k}$).  Suppose we start with the same previous estimate and error covariance matrix for both filters.

\begin{equation} \label{x0}
\hat{x}^D_{k-1|k-1} = \hat{x}_{k-1|k-1}
\end{equation}

\begin{equation} \label{P0}
P^D_{k-1|k-1} = P_{k-1|k-1}
\end{equation}

Thus, we consider the benefit of using the new constrained filter over the unconstrained Kalman Filter gained in one iteration.  We can re-write all the constrained filter's equations in terms of the corresponding equations of the unconstrained Kalman Filter.

Starting with Equation \eqref{kfspd}, we find that the state prediction remains the same over one iteration.

\begin{subequations}
\begin{align}
\hat{x}^D_{k|k-1} &\stackrel{\eqref{x0}}{=} F_{k,k-1} \hat{x}_{k-1|k-1} \\
&\stackrel{\eqref{kfsp}}{=} 
	\hat{x}_{k|k-1}
\end{align}
\end{subequations}

Similarly, we find the error covariance prediction from Equation \eqref{kfcpd} remains the same over one iteration.

\begin{subequations}
\begin{align}
P^D_{k|k-1} &\stackrel{\eqref{P0}}{=} F_{k,k-1} P_{k-1|k-1} F_{k,k-1}' + Q_{k,k-1} \\
&\stackrel{\eqref{kfcp}}{=} 
	P_{k|k-1}
\end{align}
\end{subequations}


The measurement prediction from Equation \eqref{kfmpd} is then modified as below.

\begin{subequations}
\begin{align} 
\hat{z}^D_{k|k-1} & \stackrel{\eqref{x0}}{=}
	\begin{bmatrix}
		H_{k} \hat{x}_{k|k-1}\\
		D_k \hat{x}_{k|k-1}
	\end{bmatrix} \\
& \stackrel{\eqref{kfmp}}{=}
	\begin{bmatrix}
		\hat{z}_{k|k-1}\\
		D_k \hat{x}_{k|k-1}
	\end{bmatrix}
\end{align}
\end{subequations}

We can also easily modify the measurement residual from Equation \eqref{kfid}.

\begin{subequations} \label{kfid2}
\begin{align} 
\nu^D_{k} & \stackrel{\eqref{x0}}{=}
	\begin{bmatrix}
		z_k - H_{k} \hat{x}_{k|k-1}\\
		\delta_k -D_k \hat{x}_{k|k-1}
	\end{bmatrix}\\
& \stackrel{\eqref{kfi}}{=} 
	\begin{bmatrix}
		\nu_k\\
		\delta_k -D_k \hat{x}_{k|k-1}
	\end{bmatrix}
\end{align}
\end{subequations}

And the measurement residual covariance from Equation \eqref{kficd} can then be modified as well.

\begin{subequations} \label{S_reduced}
\begin{align} 
S^D_{k} & \stackrel{\eqref{P0}}{=} 
	\begin{bmatrix}
		H_k P_{k|k-1} H_k' + R_k & H_k P_{k|k-1} D_k' \\
		D_k P_{k|k-1} H_k' & D_k P_{k|k-1} D_k'
	\end{bmatrix} \\
& \stackrel{\eqref{kfic}}{=} 
	\begin{bmatrix}
		S_k & H_k P_{k|k-1} D_k' \\
		D_k P_{k|k-1} H_k' & D_k P_{k|k-1} D_k'
	\end{bmatrix}
\end{align}
\end{subequations}

As before, we are interested in finding $\left(S_k^D\right)^{-1}$ in a block structure.  We follow the methodology described in Appendix \ref{sec::analytical Sinv} and apply it to Equation \eqref{S_reduced}.\footnote{When finding $\left(S_k^D\right)^{-1}$  as described above, we know that $A$ as defined in Appendix \ref{sec::analytical Sinv} will be nonsingular since it represents the measurement residual covariance $S_k$.  If this matrix was singular, this would mean there exists no uncertainty in our measurement prediction {\em or} in our measurement, and thus there would be no ability to filter.  Similarly, we know that $J$ as defined in Appendix \ref{sec::analytical Sinv} must also be nonsingular, which is equal to $D_k P_{k|k-1} D_k'$ (see Equation \eqref{S_reduced}).  This term projects the predicted error covariance down to the constrained space.  For well defined constraints (as described earlier), this will never be singular -- it will have the same rank as $D_k$.}

\begin{subequations} \label{SDinv_a_simp}
\begin{align}
\left(S_k^D\right)^{-1}_a  & \stackrel{\eqref{kfic}}{=} 
	&& \nonumber S_k^{-1} + S_k^{-1} H_k P_{k|k-1} D_k' \\
	&&& \nonumber \left(D_k P_{k|k-1} D_k' - D_k P_{k|k-1} H_k' S_k^{-1} H_k \right.\\
	&&& \left. P_{k|k-1} D_k'\right)^{-1} D_k P_{k|k-1} H_k' S_k^{-1} \\
& \stackrel{\eqref{DPD}}{=}
	&& \nonumber S_k^{-1} + S_k^{-1} H_k P_{k|k-1} D_k' \\
	&&&\left(D_k P_{k|k} D_k'\right)^{-1} D_k P_{k|k-1} H_k' S_k^{-1} \\
& \stackrel{\eqref{KDDPDDK}}{=}
	&& S_k^{-1} + K_k' D_k' \left(D_k P_{k|k} D_k'\right)^{-1} D_k K_k
\end{align}
\end{subequations}

In a similar manner using Equations \eqref{kfic}, \eqref{DPD}, and \eqref{KDDPDDK}, we arrive at the following remaining terms in the block structure.

\begin{align} \label{SDinv_b_simp}
\left(S_k^D\right)^{-1}_b  = & - K_k' D_k' \left(D_k P_{k|k} D_k'\right)^{-1} 
\end{align}

\begin{align} \label{SDinv_c_simp}
\left(S_k^D\right)^{-1}_c  = & - \left(D_k P_{k|k} D_k' \right)^{-1} D_k K_k
\end{align}

\begin{align} \label{SDinv_d_simp}
\left(S_k^D\right)^{-1}_d  = & \left(D_k P_{k|k} D_k' \right)^{-1}
\end{align}


Applying this to Equations \eqref{KDa}, we can find the first part of the Kalman Gain.


\begin{subequations} \label{KDa_simp}
\begin{align}
\left(K_k^D\right)_a & \stackrel{\eqref{P0}}{=}  
	&& \nonumber P_{k|k-1} H_k' \left(S_k^D\right)^{-1}_a \\
	&&& + P_{k|k-1} D_k' \left(S_k^D\right)^{-1}_c \\
&\stackrel{\eqref{SDinv_a_simp},\eqref{SDinv_c_simp}}{=} 
	&& \nonumber P_{k|k-1} H_k' S_k^{-1} \\
	&&& \nonumber + P_{k|k-1} H_k' K_k' D_k' \left(D_k P_{k|k} D_k' \right)^{-1}\\
	&&& \nonumber \qquad D_k K_k \\
	&&&  - P_{k|k-1} D_k' \left(D_k P_{k|k} D_k' \right)^{-1} D_k K_k \\
&\stackrel{\eqref{kfkg}}{=} 
	&& K_k - \left(P_{k|k-1} - P_{k|k-1}  H_k' K_k' \right) \\
	&&& D_k' \left(D_k P_{k|k} D_k' \right)^{-1} D_k K_k \\
&\stackrel{\eqref{P-PHK}}{=} 
	&& K_k - P_{k|k} D_k' \left(D_k P_{k|k} D_k' \right)^{-1} D_k K_k 	
\end{align}
\end{subequations}

Following similar steps using Equations \eqref{P0}, \eqref{SDinv_b_simp}, \eqref{SDinv_d_simp}, and \eqref{P-PHK}, we can arrive at the other part of the Kalman Gain.

\begin{equation} \label{KDb_simp}
\left(K_k^D\right)_b = P_{k|k} D_k' \left(D_k P_{k|k} D_k'\right)^{-1}
\end{equation}

We can then substitute our expressions for $K^D_k$ directly into Equation \eqref{kfsud} to find a simplified form of the updated state estimate.

\begin{subequations} \label{ckfx}
\begin{align}
\hat{x}^D_{k|k} & \stackrel{\eqref{x0}}{=}  && \hat{x}_{k|k-1} +K^D_{k}  \nu^D_{k} \\
&\stackrel{\eqref{kfkgd},\eqref{kfid2}}{=}&&
	\nonumber \hat{x}_{k|k-1} + \left(K_k^D\right)_a \nu_k\\
	&&& + \left(K_k^D\right)_b \left(\delta_k -D_k \hat{x}_{k|k-1}\right) \\
&\stackrel{\eqref{KDa_simp},\eqref{KDb_simp}}{=}
	&& \nonumber \hat{x}_{k|k-1} + K_k \nu_k \\
	&&& \nonumber - P_{k|k} D_k' \left(D_k P_{k|k} D_k' \right)^{-1} D_k K_k \nu_k \\
	&&& \nonumber + P_{k|k} D_k' \left(D_k P_{k|k} D_k'\right)^{-1} \\
	&&& \qquad \left(\delta_k -D_k \hat{x}_{k|k-1} \right) \\
&\stackrel{\eqref{kfsu}}{=}&& \nonumber
	\hat{x}_{k|k} - P_{k|k} D_k' \left(D_k P_{k|k} D_k' \right)^{-1} D_k \\
	&&& \nonumber \qquad \left( \hat{x}_{k|k} - \hat{x}_{k|k-1}\right) \\
	&&& \nonumber + P_{k|k} D_k' \left(D_k P_{k|k} D_k'\right)^{-1} \\
	&&& \qquad \left(\delta_k -D_k \hat{x}_{k|k-1} \right) \\
&=&&
	\nonumber \hat{x}_{k|k} - P_{k|k} D_k' \left(D_k P_{k|k} D_k' \right)^{-1} \\
	&&& \qquad \left(D_k \hat{x}_{k|k} - \delta_k \right)
\end{align}
\end{subequations}

Similarly, we can expand the updated error covariance in Equation \eqref{kfcud}.

\begin{subequations} \label{ckfp}
\begin{align} 
\label{ckfcua}P^D_{k|k} & \stackrel{\eqref{P0}}{=}  && \left(I - K^D_{k} H^D_{k}\right) P_{k|k-1} \\
& \stackrel{\eqref{kfkgd},\eqref{HkD}}{=} && 
	\left(I -  \left(K_k^D\right)_a H_k - \left(K_k^D\right)_b D_{k}\right) P_{k|k-1} \\
& \stackrel{\eqref{KDa_simp},\eqref{KDb_simp}}{=} &&
	\nonumber \left(I - K_k H_k + P_{k|k} D_k' \left(D_k P_{k|k} D_k' \right)^{-1} D_k \right. \\
	&&& \nonumber \left. K_k H_k - P_{k|k} D_k' \left(D_k P_{k|k} D_k'\right)^{-1} D_k \right) \\
	&&& P_{k|k-1} \\
& = &&
	\nonumber \left(I - K_k H_k \right) P_{k|k-1} - P_{k|k} D_k' \\
	&&&  \left(D_k P_{k|k} D_k' \right)^{-1} D_k \left(I - K_k H_k \right)P_{k|k-1} \\
& \stackrel{\eqref{kfcu}}{=} && 
	P_{k|k} -  P_{k|k} D_k' \left(D_k P_{k|k} D_k' \right)^{-1} D_k P_{k|k}
\end{align}
\end{subequations}

Equations \eqref{ckfx} and \eqref{ckfp} give us the improvement gained over an unconstrained Kalman Filter in a single iteration of the augmentation approach to constrained Kalman Filtering.  We see that the covariance matrix can only get tighter since we are subtracting a positive semi-definite matrix from $P_{k|k}$ above.

\section{Incorporating Equality Constraints by Projecting the Unconstrained Estimate} \label{sec::pue}

The second approach to equality constrained Kalman Fitlering is to run an unconstrained Kalman Filter and to project the estimate down to the constrained space at each iteration.  We can then feed the new constrained estimate into the unconstrained Kalman Filter and continue this process.  Such a method can be described by the following minimization problem for a given time-step $k$, where $\hat{x}_{k|k}^P$ is the constrained estimate,  $\hat{x}_{k|k}$ is the unconstrained estimate from the Kalman Filter equations, and $W_k$ is any positive definite symmetric weighting matrix.

\begin{equation}
\hat{x}_{k|k}^P = \arg\min_{x} \left\{\left(x - \hat{x}_{k|k} \right)' W_k \left(x - \hat{x}_{k|k} \right) : D_k x = \delta_k\right\}
\end{equation}

The best constrained estimate is then given by

\begin{equation}
\hat{x}_{k|k}^P = \hat{x}_{k|k} - W_k^{-1} D_k' \left( D_k W_k^{-1} D_k' \right)^{-1} \left(D_k \hat{x}_{k|k} - \delta_k \right)
\end{equation}

If we choose $W_k = P_{k|k}^{-1}$, we obtain the same solution as Equation \eqref{ckfx}.  This is not obvious considering the differing approaches.  The updated error covariance under this assumption will be the same as Equation \eqref{ckfp} since $P_{k|k}^P = \mathbb{E}\left[\left(x_k - \hat{x}_{k|k}^P\right)\left(x_k - \hat{x}_{k|k}^P\right)'\right]$ and $\hat{x}_{k|k}^P =  \hat{x}_{k|k}^D$.  Further this choice of $W_k$ is the most natural since it best describes the uncertainty in the state.  

\section{Dealing with Nonlinearities} \label{sec::nl}

Thus far, in the Kalman Filter we have dealt with linear models and constraints.  A number of methods have been proposed to handle nonlinear constraints.  In this paper, we will focus on the most widely known of these, the Extended Kalman Filter.  Let's re-write the discrete unconstrained Kalman Filtering problem from Equations \eqref{kfsm} and \eqref{kfmm} below, incorporating nonlinear models.

\begin{equation} \label{kfsm-nl} 
x_{k} = f_{k,k-1} \left(x_{k-1}\right) + u_{k,k-1}, \quad u_{k,k-1} \sim N(0,Q_{k,k-1}) 
\end{equation}

\begin{equation} \label{kfmm-nl} 
z_{k} = h_{k} \left(x_{k}\right) + v_{k}, \quad v_{k} \sim N(0,R_{k}) 
\end{equation}

In the above equations, we see that the transition matrix $F_{k,k-1}$ has been replaced by the nonlinear vector-valued function$f_{k,k-1}\left(\cdot\right)$, and similarly, the matrix $H_k$, which transforms a vector from the state space into the measurement space, has been replaced by the nonlinear vector-valued function $h_k\left(\cdot\right)$.  The method proposed by the Extended Kalman Filter is to linearize the nonlinearities about the current state prediction (or estimate).  That is, we choose $F_{k,k-1}$ as the Jacobian of $f_{k,k-1}$ evaluated at $\hat{x}_{k-1|k-1}$, and $H_k$ as the Jacobian of $h_k$ evaluated at $\hat{x}_{k|k-1}$ and proceed as in the linear Kalman Filter of Section \ref{sec::kf}.\footnote{We can also do a midpoint approximation to find $F_{k,k-1}$ by evaluating the Jacobian at $\hat{x}_{k-1|k-1}$ and $\hat{x}_{k|k-1}$ and then taking the component-wise mean.  This has the disadvantage that it is twice as expensive for finding $F_{k,k-1}$, but it should be a much closer approximation.  We use this approximation for the Extended Kalman Filter example later in this paper.}  Numerical accuracy of these methods tends to depend heavily on the nonlinear functions.  If we have linear equality constraints but a nonlinear $f_{k,k-1}\left(\cdot\right)$ and $h_k\left(\cdot\right)$, we can adapt the Extended Kalman Filter to fit into the framework of the methods described in Sections \ref{sec::ams} and \ref{sec::pue}.  We have chosen to omit the specific equations, as the extension should be apparent.

\subsection{Nonlinear Equality Constraints}

Since equality constraints we model are often times nonlinear, it is important to make an extension to nonlinear equality constrained Kalman Filtering for the two methods discussed thus far.  We replace the linear equality constraint on the state space by the following nonlinear constraint $d_k\left(x_k\right) = \delta_k$, where $d_k\left(\cdot\right)$ is a vector-valued function.  The method based on augmenting the constraints presented in Section \ref{sec::ams} is trivially extended by using an Extended Kalman Filter before -- i.e., we choose $D_k$ in Equation \eqref{HkD} as the Jacobian of $d_k$ evaluated at $\hat{x}_{k|k-1}^D$.

Incorporating nonlinear equality constraints into the projection method described in Section \ref{sec::pue} requires a more explicit change.  If we linearize our constraint, $d_k\left(x_k\right) = \delta_k$, about the current state prediction $\hat{x}_{k|k-1}$, we have the following.

\begin{equation}
d_k\left(\hat{x}_{k|k-1}^P\right) + D_k \left(x_k - \hat{x}_{k|k-1}^P \right) \approx \delta_k
\end{equation}

Here $D_k$ is defined as the Jacobian of $d_k$ evaluated at $\hat{x}_{k|k-1}^P$, similar to before.  This indicates then, that the nonlinear constraint we would like to model can be approximated by the following linear constraint

\begin{equation} \label{puenl}
D_k x_k \approx \delta_k + D_k \hat{x}_{k|k-1}^P - d_k\left(\hat{x}_{k|k-1}^P\right)
\end{equation}

Then our projected state is given as in Section \ref{sec::pue}, with $D_k$ defined as above, and $\delta_k$ replaced by the right hand side of Equation \eqref{puenl}.

\section{Discussion of Methods}

Thus far, we have discussed two different methods for incorporating equality constraints in a Kalman Filter, and we have shown that both are mathematically equivalent under the assumption that the weighting matrix $W_k$ chosen in Section \ref{sec::pue} is chosen to be $P_{k|k}^{-1}$.  As such, the projection method is a more general formulation of the augmentation method described in Section \ref{sec::ams}.  On the other hand, the augmentation method provides a trivial extension to {\em soft} equality constrained Kalman Filtering by increasing the noise modeled in $R_k^D$ to reflect how soft the constraint should be.

In implementations, there are some subtle differences.  For instance, the first method requires a minimal adjustment to codes for an existing Kalman Filter or an Extended Kalman Filter -- i.e., we can pass in the augmented matrices and get the constrained estimate.  This is especially advantageous for codes that use variations of the standard linear Kalman Filter (e.g., an Unscented Kalman Filter).  On the other hand, the second method will require less memory and computation, which can significantly speed up the filtering when the state space and constraint space are both large.  The second method does not store or compute the `cross-correlation' terms of Equation \eqref{S_reduced}, which are most likely of little interest to the engineer.

There is another more transparent difference between these two methods.  In implementations, we are bound to receive numerical round-off error.  While these two methods are mathematically equivalent, we will not see the exact same result.  The round off error that causes the most problem occurs when the updated error covariance $P_{k|k}^D$ or $P_{k|k}^P$ lose symmetry or positive definiteness.  A way around this is to use the Joseph Form of the updated error covariance (see \cite{BLK2001}) -- this will be discussed further in another publication.

\section{Conclusions}

We've presented two approaches for incorporating state space equality constraints into a Kalman Filter and shown that both result in the same estimate structure under certain conditions. The projection method should prove to be computationally faster and is also a generalization that allows different weighting matrices when projecting the estimate.  However, the augmentation method may prove easier in implementations since we can use an existing Kalman Filter without any code modifications.  We can also easily extend the latter to enforce {\em soft} equality constraints, where we allow the constraint to be slightly blurred by adding a proportionate amount of noise $R_d^K$ (Equation \eqref{RD}).
\appendix

\section{An analytic block representation for $\left(S_k^D\right)^{-1}$} \label{sec::analytical Sinv}

$S^D_k$ as defined in Equation \eqref{kficd} is a symmetric saddle point matrix of the form $M_{SPM}$ below.

\begin{equation} \label{spm}
M_{SPM} =
	\begin{bmatrix}
		A & B' \\
		B & -C
	\end{bmatrix}
\end{equation}

In the case that $A$ is nonsingular and the Schur complement $J = -\left(C + B A^{-1} B'\right)$ is also nonsingular in the above equation, it is known that the inverse of this saddle point matrix can be expressed analytically by the following equation (see e.g., \cite{BGL2005}).

\begin{equation}
M_{SPM}^{-1} =
	\begin{bmatrix}
		A^{-1} + A^{-1} B'  J^{-1} B A^{-1} & -A^{-1} B' J^{-1} \\
		-J^{-1} B A^{-1} & J^{-1}
	\end{bmatrix}
\end{equation}

For $S^D_k$, we have the following equations to fit the block structure of Equation \eqref{spm} (see Equation \eqref{kficd}).  

\begin{align}
A & =  H_k P^D_{k|k-1} H_k' + R_k \\
B & = D_k P^D_{k|k-1} H_k' \\
C & = -D_k P^D_{k|k-1} D_k'
\end{align}

Under the assumption that both $A$ and $J$ are nonsingular, we can make some substitutions and express $\left(S_k^D\right)^{-1}$ following the notation of Equation \eqref{Sinv}.





\section{Some Identities}

The following are identities that will prove useful in some of the earlier derivations of Section \ref{sec::ams}.  The matrices in these identities are used as defined in Sections \ref{sec::kf} and \ref{sec::ams}.

\subsection{First Identity}

\begin{subequations} \label{DPD}
\begin{align}
D_k & P_{k|k-1} D_k' - D_k P_{k|k-1} H_k' S_k^{-1} H_k P_{k|k-1} D_k' \\
&\stackrel{\eqref{kfkg}}{=} 
	D_k P_{k|k-1} D_k' - D_k K_k H_k P_{k|k-1} D_k' \\
&= 
	D_k \left(I - K_k H_k \right) P_{k|k-1} D_k' \\
&\stackrel{\eqref{kfcu}}{=} 
	D_k P_{k|k} D_k'
\end{align}
\end{subequations}

\subsection{Second Identity}

In the first step below, we make use of the symmetry of $P_{k|k-1}$ and $S_k^{-1}$.

\begin{subequations} \label{KDDPDDK}
\begin{align}
S_k^{-1}&&&\hspace{-.35in}H_k P_{k|k-1} D_k' \left(D_k P_{k|k} D_k'\right)^{-1} D_k P_{k|k-1} H_k' S_k^{-1} \\
& = &&
	 \nonumber \left(P_{k|k-1} H_k' S_k^{-1} \right)' D_k' \left(D_k P_{k|k} D_k'\right)^{-1} \\
	 &&&  D_k P_{k|k-1} H_k' S_k^{-1}  \\
&\stackrel{\eqref{kfkg}}{=} &&
	K_k' D_k' \left(D_k P_{k|k} D_k'\right)^{-1} D_k K_k 
\end{align}
\end{subequations}

\subsection{Third Identity}

\begin{subequations}  \label{P-PHK}
\begin{align}
P_{k|k-1} & - P_{k|k-1} H_k' K_k'\\
& \hspace{-.2in}= \label{idsym1}
	P_{k|k-1} \left(I - H_k'K_k' \right) \\
& \hspace{-.2in}= \label{idsym2}
	\left(I - K_k H_k \right) P_{k|k-1} \\
& \hspace{-.2in}\stackrel{\eqref{kfcu}}{=}
	P_{k|k}
\end{align}
\end{subequations}

Again, we've made use of the symmetry of $P_{k|k-1}$ between Equations \eqref{idsym1} and \eqref{idsym2}.

\bibliography{ccc}

\end{document}